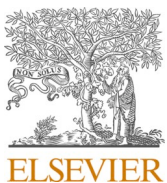
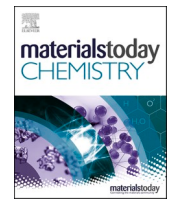

# Modulating the magnetic properties of Fe$_3$C/C encapsulated core/shell nanoparticles for potential prospects in biomedicine

A. Castellano-Soria [a,b,*], R. López-Méndez [c], A. Espinosa [d,c], C. Granados-Miralles [e], M. Varela [b,f], P. Marín [a,b], E. Navarro [a,b], J. López-Sánchez [e]

[a] *Instituto de Magnetismo Aplicado (IMA−UCM−ADIF), 28230, Madrid, Spain*
[b] *Departamento de Física de Materiales, Facultad de Físicas, Universidad Complutense de Madrid (UCM), 28040, Madrid, Spain*
[c] *IMDEA Nanociencia, c/ Faraday, 9, 28049, Madrid, Spain*
[d] *Instituto de Ciencia de Materiales de Madrid (ICMM−CSIC), 28049, Madrid, Spain*
[e] *Instituto de Cerámica y Vidrio (ICV−CSIC), 28049, Madrid, Spain*
[f] *Instituto Pluridisciplinar, Universidad Complutense de Madrid (UCM), 28040, Madrid, Spain*

ABSTRACT

In the pursuit of alternative and less invasive medical treatments, magnetic nanoparticles (NPs) have gained significant relevance. Iron carbides NPs stand out for their higher saturation magnetizations compared to iron oxides, while maintaining a suitable biocompatibility. In this work, high control is achieved over the composition and morphology of Fe$_3$C/C encapsulated core/shell nanoparticles through fine-tuning of the sol-gel synthesis parameters. Specifically, the impact of decreasing each surfactant concentration added, $n_t$, the same both for oleylamine (ON) and oleic acid (OA), has been explored. A minimum value for such parameter denoted by $n_{t,min.}$ was required to produce pure Fe$_3$C@C NP-composites. For $n_t < 4$ mmol, some minor α-Fe impurities arise, and the effective carburization becomes unstable due to insufficient carbon. The magnetic properties of the materials prepared were optimized by reducing the excess carbon from surfactants, resulting in saturation magnetization values of 86 emu/g. (for pure Fe$_3$C at $n_t = 5$ mmol) and 102 emu/g (for Fe$_3$C and <2 % w.t. of α-Fe impurity at $n_t = 4$ mmol). In view of this, several cytotoxicity studies for different Fe$_3$C@C samples were conducted, exhibiting excellent biocompatibility in cell-based assays, which could lead to potential application at the forefront of biomedical fields.

## 1. Introduction

Within the current landscape of nanostructured magnetic materials, metal-carbon based materials have emerged as a focal point for extensive research and exploration [1–3]. Magnetic Fe-carbide nanoparticles (NPs) have aroused interest due to the combination of functional properties induced by the presence of carbon, providing a high chemical stability against oxidation [4], improved catalytic properties [5,6] and remarkable biocompatibility [7]. Although Fe-oxide NPs have been widely studied for biomedical applications due to the tunability of their magnetic properties, low toxicity, and biodegradability [8,9], their application in biomedicine is sometimes limited as a consequence of the high saturation magnetization ($M_s$) values required to boost the theragnostic effect. In this context, Fe-carbides stand out with increased $M_s$ ~140 emu/g for both θ-Fe$_3$C [10–13] and χ-Fe$_5$C$_2$ [14], and ~160 emu/g for εFe$_2$C [14], compared to the 30–50 % lower values of Fe-oxides: 80 and 92 emu/g for γ-Fe$_2$O$_3$ and Fe$_3$O$_4$ respectively [15]. This fact makes them good candidates for future biomedical applications [16], in particular, for those related to magnetic-dependent antitumor theragnostic purposes [17–19]. The few published hyperthermia studies based on Fe$_3$C and Fe$_{2.2}$C NPs [20–23] have provided interesting results with specific absorption rate (SAR) values ranging from 80 to 1700 W/g, competitive with those typically reported for Fe-oxides. Moreover, the use of Fe$_5$C$_2$ NPs has shown an efficient performance both as contrast agent for magnetic resonance imaging (MRI) [24,25] and in photothermal cancer therapy studies [26]. Although these results are limited compared to the extensively explored paradigm of Fe-oxides [27,28], the preliminary results are encouraging. In addition, some recent studies have shown that Fe-carbides have an even lower natural cytotoxicity than Fe-oxides [29]. For instance, chemically modified carbon-coated iron NPs (Fe/C NP composites) have exhibited high biocompatibility [30], and demonstrated potential anti-atherosclerotic plaque properties [31,32]. The $Fe_3C$–C–N dopped NP-systems have shown excellent biocompatibility at concentrations close to 300 μg mL$^{-1}$ with a cell survival rate of 80 % and a good peroxidase activity, catalyzing $H_2O_2$ to OH radicals and boosting the antibacterial action of $H_2O_2$ [33]. Long-term *in vivo* studies have also been conducted in mice for incubation periods of one week and one year of $Fe_3C$/C core/shell NPs, with exceptionally low toxicity [34]. Similarly, in phototoxicity experiments the $Fe/Fe_3C$ NPs revealed a significant cell death rate of nearly 80 % for concentrations that showed low cytotoxicity results (95–100 % of viability), ranging from 75 to 200 μg mL$^{-1}$ [35]. Also, other iron carbide phases, such as $Fe_7C_3$, showed high viability after 24 h of incubation and no effect on the cytophysiological parameters of in vitro cultured cells [36] such as mitotic division or DNA replication processes.

In view of these preliminary studies, a great deal of efforts has been directed towards designing synthesis procedures which allow reproducibly obtaining single phase $Fe_3C$ (cementite, *Pnma* 62) NPs with good particle size control, as this is the most hemodynamically stable of the metastable phases of iron carbides [37]. In the framework of the various successful synthesis routes for $Fe_3C$ NPs documented including colloidal wet-chemistry synthesis [38–40], plasma discharge [41], nanocasting [42], metalorganic chemical vapor deposition [11], glucose pyrolysis [43] or modified sol-gel routes [10,44,45], we recently reported a novel modified macromolecule-sol-gel chemical route for the preparation of single phase $Fe_3C$/C-graphite core/shell nanoparticles [13]. The highly effective Fe-carburization of primal Fe-oxide NPs (formed at T < 600°) by the aliphatic carbon of the macromolecules selected as surfactants, when the xerogel is pyrolyzed at temperatures of 600–800 °C, resulted in a high magnetic response together with a high chemical stability of the NPs obtained.

In this work, we exploit the above-mentioned novel synthesis for the preparation of different xerogels with the aim of optimizing the magnetic response of the NPs, desirable for biomedical applications. For a fixed amount of Fe added into the synthesis, different amounts of surfactants were tested to determine the threshold of amorphous-carbon matrix content at which the nanoparticles are $Fe_3C$ single phase after pyrolyzing the xerogels at a fixed temperature of 700 °C. This procedure allows to modulate both the saturation magnetization and the coercivity field ($H_C$) of the $Fe_3C$@C NPs systems achieving an $M_s$ up to 102 emu/g and a semihard character ($H_C \sim 600$ Oe) as the carbon matrix content approaches the threshold value. Morphological and phase analysis of the NPs were also conducted to assess the impact of decreasing the total surfactant concentration and its influence on particle coalescence and compositional phase modulation, the two last being tightly correlated to the magnetic properties. The limiting amount of surfactant is experimentally determined as the required to reduce the primal Fe-oxide NPs (formed at T < 700 °C) and efficiently carburize them to $Fe_3C$ (at 700 °C). Moreover, a theoretical model for predicting the final carbon matrix content after pyrolyzing the xerogels (700 °C) was introduced, achieving a good correlation with the experimental values and with the potential to be extended to other sol-gel synthesis using macromolecules as surfactants. In addition, a review of the $Fe_3C$ NPs state-of-the-art is presented, which shows that the magnetic values of the samples prepared here are competitive with those reported in the literature, highlighting the high degree of purity degree of the $Fe_3C$ single phase, which is not always achieved. Finally, based on such surfactant threshold, several sets of $Fe_3C$/C samples were selected to assess the viability of the compounds in MCF-7 tumour cells (breast cancer cell line), showing low cytotoxicity values for an incubation time of 24 h at 37 °C and concentrations ranging from 10 to 100 μg/mL suggesting a potential applicability in biomedicine.

## 2. Experimental techniques

### 2.1. Synthesis of the $Fe_3C$/C encapsulated core/shell nanoparticles by sol-gel method

The NPs have been prepared by a novel one-pot modified sol-gel synthesis approach. This process involved surfactants with large hydrophobic tails and high boiling points, namely oleic acid (OA, 360 °C) and oleylamine (ON, 364 °C). Hydrated iron nitrate $Fe(NO_3)_3·9H_2O$ (Alfa Aesar +98 %) was used as a salt precursor. After dissolving $n_s = 12$ mmol of salt in absolute ethanol (PanReac +99.8 %) at 40 °C for 15 min, each surfactant was added dropwise under vigorous magnetic stirring. OA was added first, followed by ON after 15 min in a 1:1 ratio. Different solutions were prepared by varying the amounts of each surfactant concentration, $n_t = \{1, 2.5, 4, 5, 10, 20$ and $30$ mmol$\}$, accounting for a total of $2n_t$ ($n_t$ of OA and $n_t$ of ON). The solutions were stirred for 24 h to obtain a homogeneous sol. Afterwards, the temperature was subsequently raised to 75 °C for 36 h to evaporate the remains of water and alcoholic groups. The resulting xerogel was pyrolyzed at 700 °C under a $N_2$ (+99.8 %) atmosphere because of its non-oxidizing nature. The formation of nanoparticles is governed by the primal micelles formed in the xerogel by the steric action of surfactants in the solution of the hydrated salt in ethanol. The hydrophilic heads of the macromolecules used as surfactants create some initial micelles where the hydrolysis and polycondensation of the cations of the hydrated salt takes place. The volume of these primal micelles is therefore limited by the hydrophilic heads and hydrophobic tails of the macromolecules used. The high boiling points of the macromolecules used lead to the crystallization of iron oxide nanoparticles at low temperatures. These nanoparticles are subsequently reduced by the remaining aliphatic carbon, which is thermally decomposed during the pyrolysis of the xerogel. The densification process was carried out at a rate of 5 °C/min for 1 h. The as-prepared materials obtained were milled for 10 min using an agate mortar. They were easily homogenized due to their brittleness, followed by two sonicated/washed cycles to eliminate the larger grains by decantation. Fig. 1 shows optical micrographs of various xerogels with different surfactant concentrations, along the resulting powder after densification and agate-milling processes.

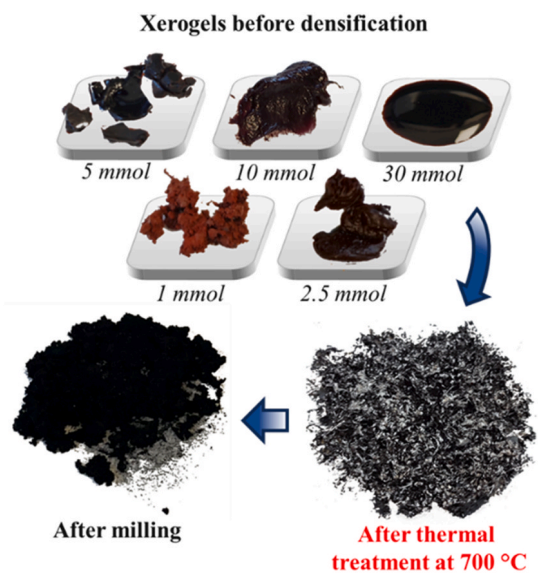

**Fig. 1.** Optical images of the xerogels prepared with different surfactant concentrations ($n_t = 1$–30 mmol), along with the resulting products from densification and agate-milling processes.

## 2.2. Thermodynamic, morphological, structural, electronic, and magnetic characterization

The thermodynamic characterization of the xerogels was performed by simultaneous thermogravimetry and differential scanning calorimetry (TG-DSC) curves that were taken in a SDT-Q600 equipment from TA instruments. The crystal structure was examined by synchrotron X-ray diffraction (SXRD) measurements performed at the BM25 beamline of the European Synchrotron Radiation Facility (ESRF) in Grenoble, France. The incident X-ray radiation used was a 25 keV beam ($\lambda$ = 0.4952 Å). The powders were introduced into a spinning quartz capillary (0.5 mm diameter) and the acquisition was performed in the 2θ range 8–57.5°, with 0.0075° step range. A 2D photon-counting X-ray *MAXIPIX* detector was employed, and the data were processed with the *BINoculars* software [46,47]. The crystal structure, lattice parameters, compositional percentage of phases and volume-weighted average crystallite size of the samples were inferred from Rietveld refinements, which were performed using the Profex interface of the BGMN Rietveld software [48].

Morphological features, crystalline structure, particle size distribution (PSD), and compositional analysis of the nanostructures encapsulated in the carbon matrix were analyzed by high-resolution scanning transmission electron microscopy (STEM) combined with a simultaneous electron energy-loss spectroscopy (EELS), using a JEOL JEM ARM200cf with a cold field emission source, working with a CEOS aberration corrector operated at 200 kV and equipped with a Gatan Quantum EELS detector. To compose the elemental maps, noise was removed from the EEL spectrum images using principal component analysis [49]. EELS elemental maps were obtained by integrating 30–40 eV wide windows under the relevant edges after background subtraction using a power law fit. The morphology of the obtained powders was analyzed using a scanning electron microscope (SEM) JEOL JSM 7600 F at 15 kV of SEI. SEM images were processed using ImageJ software [50].

The magnetic properties of the nanostructured powders were measured using a vibrating sample magnetometer (VSM) coupled to a physical property measurement system (PPMS model 6000 controller, Quantum Design). The magnetic hysteresis loops were collected at room temperature with a maximum applied magnetic field of 50 kOe.

## 2.3. Cell culture and viability study

In vitro experiments were conducted using MCF-7 cells (breast cancer cells) cultured in 24-well plates in Dulbecco's Modified Eagle Medium (DMEM) supplemented with 10 % fetal bovine serum (FBS) and 2 mM l-glutamine and 1 % penicillin at 37 °C in an atmosphere with 5 % $CO_2$. Cells were incubated for 24 h with nanoparticles at increasing concentrations (from 1 to 100 $\mu g_{TOTAL}$/mL) in complete DMEM culture medium. A 10 % of Alamar Blue® solution was incubated with DMEM medium without red phenol for around 3 h. The resulting solution for each condition was transferred to a 96-well plate. Fluorescence measurements were taken using a microplate reader with a 530 nm excitation and a 590 nm emission filters. Cell viability was calculated by comparing the results with control cells.

## 3. Results and discussion

### 3.1. Thermodynamic characterization

From all the xerogels prepared a $n_t$ = 5 mmol xerogel was selected as an example to elucidate the thermodynamic processes involved during the densification process. Previously, the xerogel is consolidated and purified with a pre-treatment performed at 250 °C for 30 min. This pre-treatment does not alter the internal structure of the iron carbide precursor micelles since the high boiling point of OA and ON (~350 °C) preserves the aliphatic carbon content and allows the evaporation of water, ethanol, and other alcoholic species mainly produced during the hydrolysis and polycondensation processes.

Fig. 2 shows the simultaneous TG-DSC curves obtained for a temperature sweep under $N_2$ atmosphere performed from 25 °C to 1000 °C at a heating rate of 5 °C/min. The DCS curve was normalized to the time-dependent mass, instead of the initial value deposited in the cup for the characterization. The curve analysis leads to a division in four thermal regions (TR1-4), bounded according to the temperature intervals defined by the five thermodynamic processes (P1-5) identified (see Table 1).

Regarding TG, weight loss occurs in two steps at TR1 and TR3, corresponding with ~30 and 20 % respectively, close to other literature reports for similar systems [51]. The initial drying pre-treatment of the xerogel up to 250 °C yields a weight loss related with some remaining water evaporation, $NO_3^-$ groups, residual ethanol, and some organic matter. In turn, the DSC curve shows a stabilization at P0, as is usually reported at the beginning of DSC curves [52]. The first activated process, P1, has an endothermic peak at 410 °C with a specific enthalpy of −1.6 J/g. This corresponds to the heat absorbed by the surfactants (OA and ON) during their boiling processes, (350 and 360 °C respectively). In TR1, TG curve shows a ~25 % weight loss at P1 associated with surfactants evaporation. Moreover, such weight loss for ON and OA is found at a higher temperature range than reported for pure OA [53,54] and ON [54], and is probably due to chemically bound surfactant on the surface of the primal Fe-oxide particles formed at low temperature (<400 °C) [54]. The TR2 region is dominated by a slight exothermic peak at 480 °C (P2) with a specific enthalpy of 0.9 J/g and the weight is stabilized at around ~67 %. In the TR2, the crystallization and consolidation of the primal low-crystalline is produced, or amorphous Fe-oxides nanoparticles into mainly $Fe_3O_4$ and FeO [13,44]. Subsequently, two endothermic processes contained in TR3 (from 665 to 720 °C) are indexed as P3 and P4 with peaks at 595 and 660 °C. The specific enthalpy values, according to the peak temperatures, suggest that the activation energy associated with P3 should be higher than P4, with −1.09 J/g and −3.56 J/g, respectively. Recent studies have reported the kinetic constants values for the carbothermal-reduction of the aforementioned oxides, $k_{FeO}$ (43(1) kJ/mol) and $k_{Fe_3O_4}$ (35(4) kJ/mol) [55]. Therefore, we associate P3 with the reduction of $Fe_3O_4 \to FeO$ and P4 with $FeO \to Fe$.

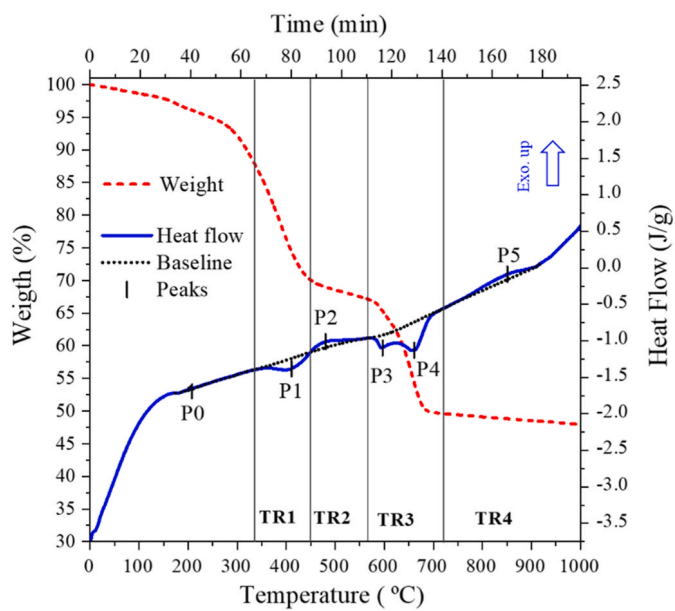

**Fig. 2.** TG (dashed-line red, left axis) and DSC (blue, right axis and exothermic criterion as "up") curves as a function of time and temperature for the 5 mmol xerogel, intending to resemble the densification step. Thermal regions denoted by TR1-4 and thermodynamic processes denoted by P0-5. (For interpretation of the references to color in this figure legend, the reader is referred to the Web version of this article.)

**Table 1**
TG-DSC results for the different processes (P1-5) ordered by thermal regions (TR1-4) and their temperature intervals. The values of temperature start-peak-end {s-p-e} $_{(P1\text{-}5)}$, and the enthalpy $H°_{(P1\text{-}5)}$ of each process is included.

|  | TR1: (335–460 °C) | TR2: (460–580 °C) | TR3: (580–710 °C) | TR4: (710–1000 °C) |
|---|---|---|---|---|
| {s-p-e} $_{(P1\text{-}5)}$, °C | {335-410-450} $_{(P1)}$ | {450-480-565} $_{(P2)}$ | {565-595-630} $_{(P3)}$ <br> {630-660-720} $_{(P4)}$ | {740-850-910} $_{(P5)}$ |
| $H°_{(P1\text{-}5)}$, (J/g) | −1.6 $_{(P1)}$ | 0.90 $_{(P2)}$ | −1.09 $_{(P3)}$ <br> −3.56 $_{(P4)}$ | 1.33 $_{(P5)}$ |

The carbon-rich environment resulting from the decomposition of organic molecules mainly into water and solid carbon, together with the Boudouard equilibrium (especially for T ≥ 700 °C, with $CO/CO_2$ ratio greater than 1) gives two possibilities for the reduction of the Fe-oxides by direct $C_{(s)}$ or $CO_{(g)}$ interaction (see Supporting Information). Even the nitrogen-rich environment can lead to the formation of highly reducing $C_xN_y$ + species at temperatures <600 °C which contribute to iron reduction [45]. Furthermore, the carburization reaction $Fe^0 + C_{(s)} \rightarrow Fe_3C_{(s)}$ occurs simultaneously and may be governed by:

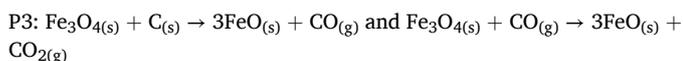

P3: $Fe_3O_{4(s)} + C_{(s)} \rightarrow 3FeO_{(s)} + CO_{(g)}$ and $Fe_3O_{4(s)} + CO_{(g)} \rightarrow 3FeO_{(s)} + CO_{2(g)}$

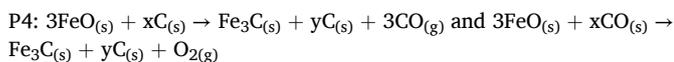

P4: $3FeO_{(s)} + xC_{(s)} \rightarrow Fe_3C_{(s)} + yC_{(s)} + 3CO_{(g)}$ and $3FeO_{(s)} + xCO_{(s)} \rightarrow Fe_3C_{(s)} + yC_{(s)} + O_{2(g)}$

Additionally, a possible transition $Fe_2O_3 \rightarrow Fe_3O_4$ may not be observed in DSC due to its fast reaction-rate and the low temperature required [55]. This transition could probably occur simultaneously overlapping with P2 or P3. Finally, after Fe-oxide reduction and its carburization, the $Fe_3C/C$ phases are consolidated in the TR3 region since carbon diffusion becomes considerable at temperatures higher than 500 °C [56]. In TR4, the broadened peak P5, shows a slow kinetics but a remarkable enthalpy of 1.33 J/g. This could be related to increased crystallization of the carbon matrix, as previously reported [13].

### 3.2. Structural properties and phase composition

To investigate the compositional variation in nanostructured materials synthesized for optimal magnetic response, several samples were prepared via densification at 700 °C. Various xerogels with different total OA and ON surfactant contents were prepared and the biocompatibility of the resultant powder samples were then evaluated. The samples were labeled according to the $n_t$ value as: 2.5, 4, 5, 10, 20 and 30 mmol (see Supporting Information).

Fig. 3-a) summarizes the diffractograms obtained for the six samples synthesized. Through Rietveld analysis compositional percentages (Fig. 3-b1) and average crystallite sizes (Fig. 3-b2) of the crystalline phases included in the refinement model can be quantitatively extracted (see Supporting Information). A predominant orthorhombic phase of the intermetallic compound $Fe_3C$ (*Pnma*, cementite) is favored in all cases, which corresponds to all the unindexed peaks of the diffractograms (Fig. 3-a). Therefore, the excess of carbon supplied by the surfactants added to the synthesis creates a favorable reducing atmosphere at elevated temperatures. From the Rietveld refinement, the lattice

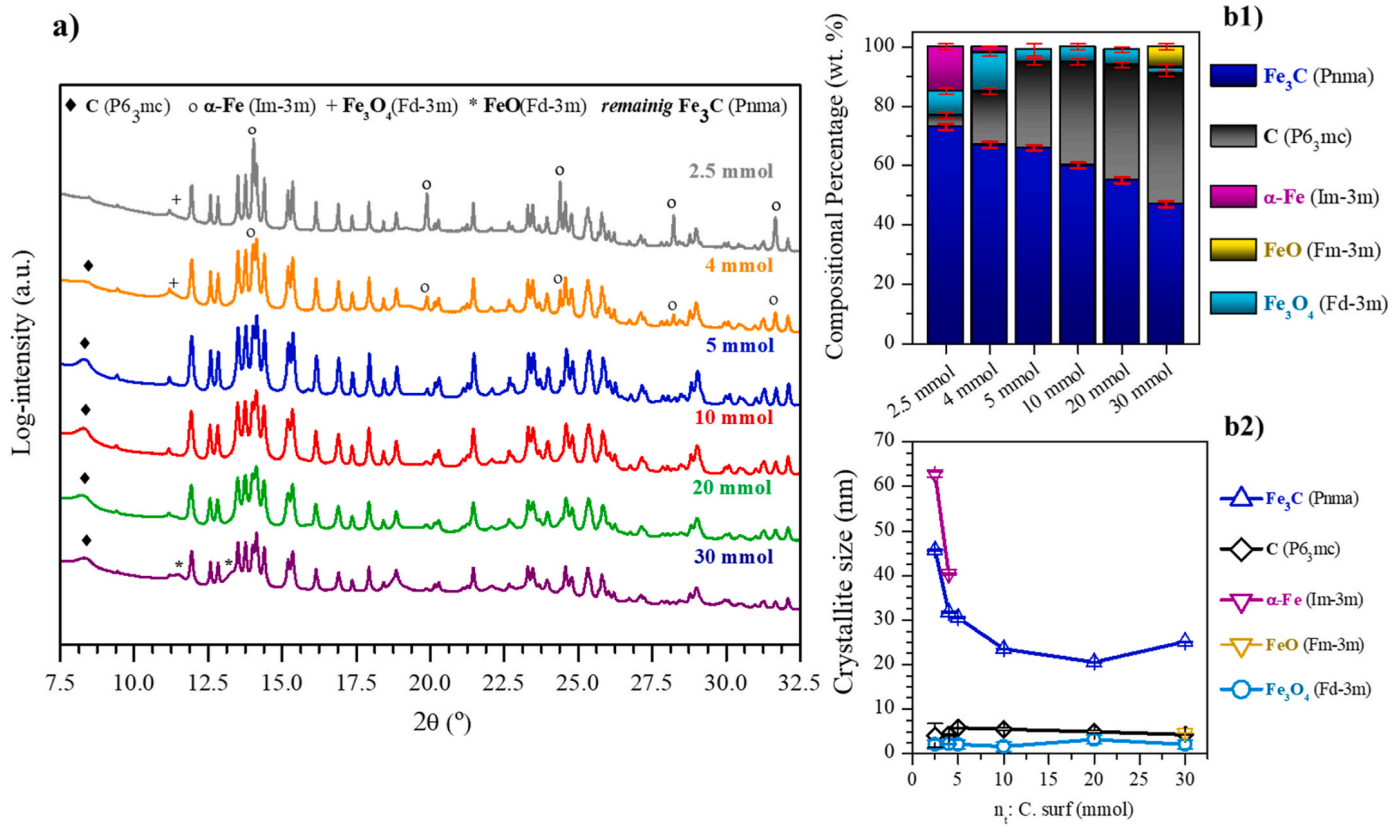

**Fig. 3.** (a) SXRD diffractograms for the samples synthesized with different surfactant concentrations ($n_t$, for each surfactant added). (b1) Mass percent composition and (b2) crystallite size of the phases calculated from Rietveld analyses.

parameters are $a_{Fe3C}$ = 5.0960(3) Å, $b_{Fe3C}$ = 6.7514(3) Å, and $c_{Fe3C}$ = 4.5321(2) Å, in good agreement with values reported by other works [12,13,57]. For the 10, 20, and 30 mmol samples, Fe$_3$C compositional percentages in Fig. 3-b1) show a practically linear decreasing trend with the increasing nt from 73 to 43 wt %, while the crystallite size in Fig. 3-b2) remains constant at about 20–25 nm. However, the 2.5, 4 and 5 mmol samples show a remarkable increase of the crystallite size up to 45 nm for the 2.5 mmol and 30 nm for the 4 and 5 mmol samples). The decrease in Fe3C compositional percentages with nt is correlated with the increase of C in the samples. In addition, all samples contain a significant amount of C (P63/mmc, graphite) which is reflected in the diffractograms by the peak 2θ = 8.25° corresponding to the Bragg (002) reflection. The carbon formed shows a lattice parameters $a_{C-g}$ = 2.451(4) Å and $c_{C-g}$ = 6.915(6) Å [58] and a roughly constant crystallite size around ~4–6 nm. The reduction in the amount of surfactant added leads to a decrease in the amount of carbon obtained in the powder after heat treatment. This decrease follows a nearly linear behavior from 30 to 5 mmol followed by an abrupt decrease for the samples 4 and 2.5 mmol. In fact, for the latter two samples, Fig. 3-a) reveals the presence of an α-Fe (Im3m) bcc phase due to an excessive reduction of carbon introduced during synthesis, causing an ineffective carburization of the reduced Fe. In this sense, previous studies have shown that 1 mmol sample is 100 % α-Fe, and no traces of C and Fe3C are found [57]. While the α-Fe compositional percentage is negligible in the 4 mmol (2.1(1)%), it increases up to 15(1)% in the 2.5 mmol sample, consolidating an α-Fe phase with the largest crystallite size (60.2(3) nm, Fig. 2 b2).

High-resolution SXRD also evidences the residual formation of two distinct Fe-oxides. The first one, Fe$_3$O$_4$ (Fd3m, spinel-like), is present in all the samples with an expanded lattice parameter of $a_{Fe3O4}$ = 8.44(2) Å due to small crystallite size ~2.5(5) nm. Its compositional percentage is low for 5–30 mmol, (~2–5%) and higher for 2.5-4 mmol (13(2) and 8(1)% respectively). Possibly, the uneven loss of surfactant in the xerogel, which happens at temperatures between 300 and 460 °C, can lead to carbon-deficient regions that are difficult to fully reduce. The second one oxide is the cubic fcc FeO (Fd3m, wüstite) which is only present in the 30 mmol sample, with a lattice parameter $a_{FeO}$ = 4.29(1) nm, compositional percentage of 7(1)% and a crystallite size of 4.6(1) nm. The C-rich environment of this xerogel would allow partial reduction of the main $Fe_3O_4$ to FeO, leaving only a 2(1)% of the spinel-like oxide.

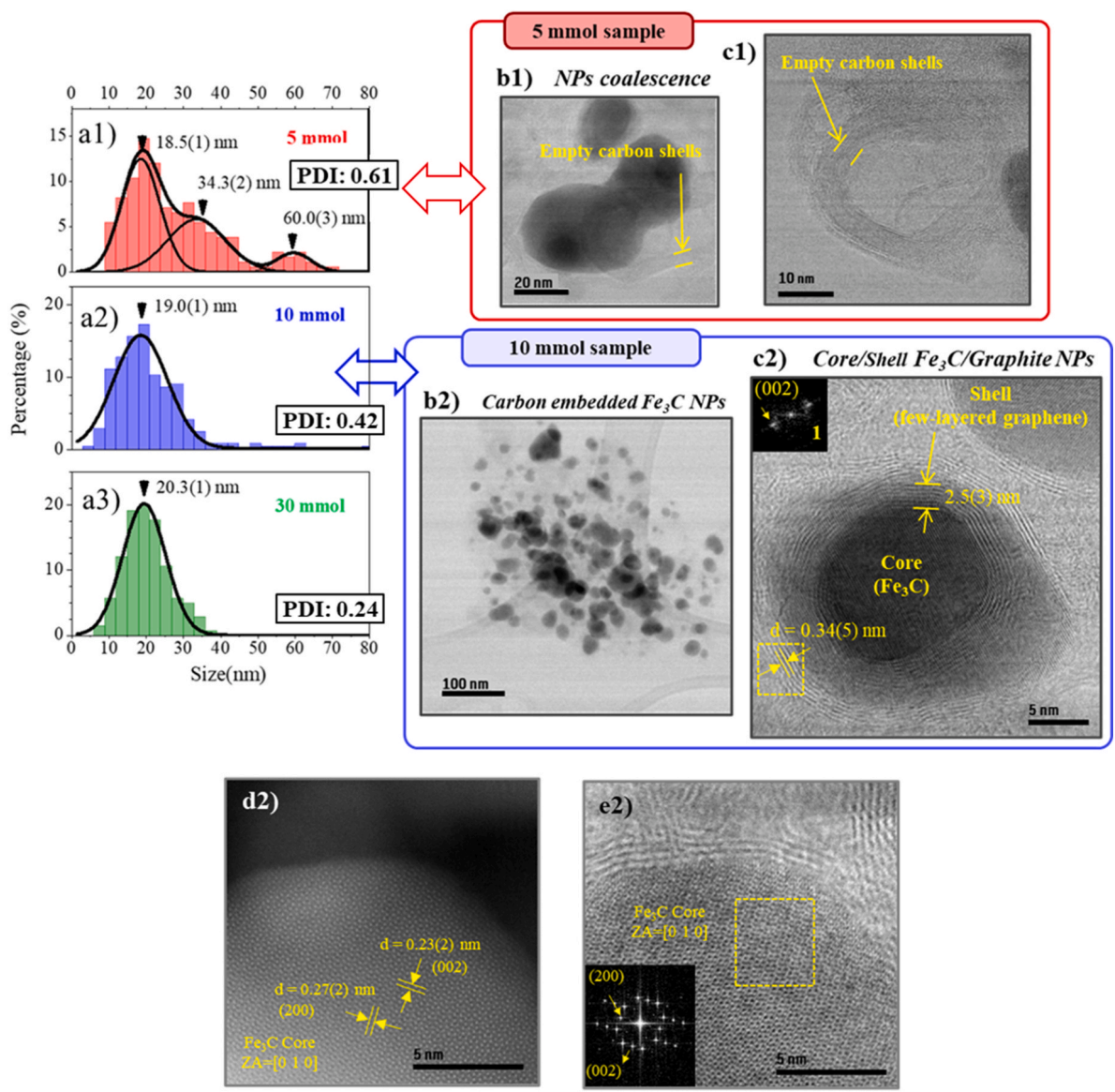

Fig. 4. (a1, a2, a3) Particle size distributions for *5, 10* and *30* mmol samples respectively with each respective PDI value. (b1, c1) Low magnification STEM ABF images of coalescence of NPs and empty carbon shells at the 5 mmol sample, respectively. (b2) Low magnification ABF image of 10 mmol Fe$_3$C NPs. (c2) ABF image of core/shell Fe$_3$C/C encapsulated NPs, and the FFT of the squared region at the inset. (d2) High resolution HAADF image of a Fe$_3$C/C NP. (e2) ABF corresponded with the (d2) image, along with the FFT of the crystalline core, such as the area within the yellow rectangle. (For interpretation of the references to color in this figure legend, the reader is referred to the Web version of this article.)

Thus, it can be concluded that due to the low values of oxides compositional percentages and the lack of α-Fe phase, the 5–30 mmol samples are almost single-phase $Fe_3C$ cores with crystalline domain sizes comprised between 20 and 30 nm. Therefore, from this range of surfactants, optimal samples will be selected for viability studies.

The composition and crystal structure of three representative samples: *5, 10* and *30* mmol were examined by STEM analyses. These samples were selected due to their low values of oxide compositional percentages and the absence of the α-Fe phase (Fig. 3). Additionally, they exhibited an approximately constant value of the $Fe_3C$ crystallite size inferred by Rietveld analysis, as demonstrated in Fig. 3-b2). Fig. 4 shows that the samples are composed of NPs embedded in a carbon matrix. Specifically, their particle size distributions (PSD) are shown in Fig. 4-a1), a2) and a3) respectively, fitted by Gaussian curves. Additionally, the polydispersity index (PDI) of each distribution is provided. The distribution for the 30 mmol sample is well described by a single curve centered at 20.3(1) nm and displays a full width high maximum (FWHM) of 13.6(1) nm with a PDI value of 0.24, close to those usually reported for NPs for biological applications [59]. For 10 mmol, the distribution becomes more asymmetric and is centered on 19.0(1) nm, reflecting a wider FWHM of 17.0(1) nm, and exhibits a value of PDI of 0.44, close to the limit of 0.5 for being considered a broad distribution [60]. In contrast, the elevated PDI value of 0.61 for the 5 mmol sample is caused by the tri-modal Gaussian distribution with three centers at 18.5 (1), 34.3(2) and 60.0(3) nm, approximately following a close-integer multiple rule by pairwise coalescence of NPs [61].

Fig. 4-b1) shows the coalescence of two $Fe_3C$ NPs favored by the lower amount of carbon matrix by reducing the amount of $n_t$ surfactants introduced in the synthesis. Fig. 4-b2) shows a low magnification annular bright field (ABF) image taken for a single grain of powder obtained for the 10 mmol sample, where $Fe_3C$ NPs are observed with spherical uneven morphology. Moreover, Fig. 4-c2) displays a single core/shell NP with $Fe_3C$ core and a well-defined multilayered graphene/graphite coating of ~2.5(3) nm thickness. It is important to note that although Rietveld analysis indicates a crystallite size of ~4–6 nm for the carbon graphite, this result may not solely represent the shells. Carbon (graphite) originating from the matrix also contributes to the XRD diffractogram. As a result, discerning between contributions from the shell and matrix associated with graphite phases may not be completely achievable. The fast Fourier transform (FFT) of the core region in Fig. 4-c2) allows to index the (002) crystallographic directions of the graphite with an associated interplanar distance of d = 0.34(5) nm and with a lattice parameter $c_{c\text{-}g}$ = 6.8(5) Å, close to that obtained by Rietveld analysis (Table 2). Thus, the strong catalytic activity of Fe and $Fe_3C$ in the formation of graphitic nanostructures or graphitization [62,63], causes the growth of the carbon shell in NPs and their chemically stabilization. In turn, Fig. 4-b,c1) show the presence of empty carbon (graphite-graphene) shells in the 5 mmol sample. This is possibly due to both the thermal agitation of the nanoparticles during the densification process and/or the insufficient carbon matrix content to constrain the thermal evolution of the NPs sizes [13], favoring the ejection of some cores to form isolated shells. In addition, a high angle annular dark field (HAADF) image of a single crystal $Fe_3C$/C NP, with a zone axis ZA =

[010] is shown in Fig. 4-d2, e2). The interplanar distances d = 0.23(2) nm and d = 0.27(2) nm corresponds to the crystallographic directions (002) and (200), respectively, as indexed by the FFT in the ABF image in Fig. 4-d2). No secondary phases are observed at the interface between the $Fe_3C$ core and the graphite shell, and the lattice parameters are consistent with those obtained from Rietveld analysis (Table 2).

Subsequently, STEM-EELS analyses are performed in the samples to quantitatively extract the local composition of the core/shell nanostructures, Fig. 4-a1-a3) displays images of NPs with $Fe_3C$ core sizes around ~5–10 nm and not fully reduced $Fe_{3-x}O_x$ shells. The shell thickness value, around 4.5(5) nm, is in close agreement with the crystallite sizes reported by Rietveld analysis (Fig. 3-b2). The low percentages of spinel-like Fe-oxide found in Fig. 3-b2) suggest a low contribution of this phase to the overall powder sample.

Fig. 5-b) illustrates an overlay of elemental compositional maps obtained from the $Fe_3C/Fe_{3-x}O_x$ NP, simultaneously displaying the integrated intensities of C K-edge (blue, Fig. 5-b1), Fe $L_{2,3}$-edge (red, Fig. 5-b2) and O K-edge (green, Fig. 5-b3). The compositional profiles along the yellow rod region are shown below. Fig. 5-b1) evidences the presence of a continuous carbon matrix around the NP and a small amount of C within the core region. The compositional profiles in Fig. 5, confirm an atomic ration of 3:1, which is consistent with the $Fe_3C$ intermetallic phase for the cores. Furthermore, Fig. 5-b3) shows a significant presence of O content in the shell. The approximately 3Fe:5O ratio derived from the EELS analysis, is close to the 3Fe:4O for pure $Fe_3O_4$ (magnetite) [15], suggesting a possible non-full spinel-like Fe-oxide stoichiometry for the shell ($Fe_{3-x}O_x$). Such shell is also found in the *20* and 30 mmol samples (not shown here). However, the low percentages of spinel-like Fe-oxide found in Fig. 3-b2) suggest a low contribution of this phase to the overall powder sample and confirming the high purity of individual $Fe_3C$/graphite NPs in samples.

Finally, it can be concluded from the morphological and compositional analyses that a concentration of at least 10 mmol is necessary to achieve a relatively narrow PSD of around 20 nm, which is suitable for viability studies.

### 3.3. Magnetic fine-tuning of $Fe_3C$/C core/shell nanostructures

Fig. 6-a) shows the hysteresis loops measured at 300 K for the synthesized samples as a function of the $n_t$. The hysteresis loop of a 1 mmol (*) sample prepared in a previous work [57] and consisting of pure α-Fe is included for comparison. The values of the coercive field and the saturation magnetization are summarized in Fig. 6-b) and Table 3. The renormalized saturation magnetization values ($M_S^R$) calculated by means of the Rietveld compositional percentages of Fig. 3-b1), i.e. the $M_s$ determined by considering the powder sample as a single pure $Fe_3C$ *Pnma* phase, are also shown (red data). For the normalization, $M_s$ (α-Fe-1 mmol sample) = 215 emu/g and $M_s$ ($Fe_3C$) = 140 emu/g [12] have been considered and the Fe-oxides contribution has been neglected. Notably, the $M_S^R$ Rietveld-predicted values are inferior to the expected for pure $Fe_3C$ since in this case, only the crystalline part of the carbon is considered while the amorphous contribution is excluded, yielding to an underestimation of the graphite matrix.

In Fig. 5-b), $H_C$ values seem to be stable for *4*–30 mmol samples, where the main magnetic contribution comes from the $Fe_3C$ phase, while for the *2.5* mmol sample it is reduced by half due to the soft

**Table 2**

STEM and Rietveld comparative analyses between maximum PSD and crystallite sizes, and lattice parameters of $Fe_3C$/graphite core/shell NPs.

| HR-STEM Analyses ◊ Rietveld Analyses | | | | | |
|---|---|---|---|---|---|
| Sample | Maximum of PSD (nm) ◊ Crystalline size (nm) | Lattice parameters (Å) "core/shell" | | | |
| 5 mmol | 18.5(1), 34.3(2), 60.0(3) ◊ 30.6(2) | – | | | |
| 10 mmol | 19.0(1) ◊ 23.5(1) | $Fe_3C$: | a = 5.0960(3) ◊ 5.4(2) | C: | c = 6.915(6) ◊ 6.8(5) |
| | | | c = 4.5321(2) ◊ 4.6(1) | | |
| 30 mmol | 20.3(1) ◊ 25.1(1) | – | | | |

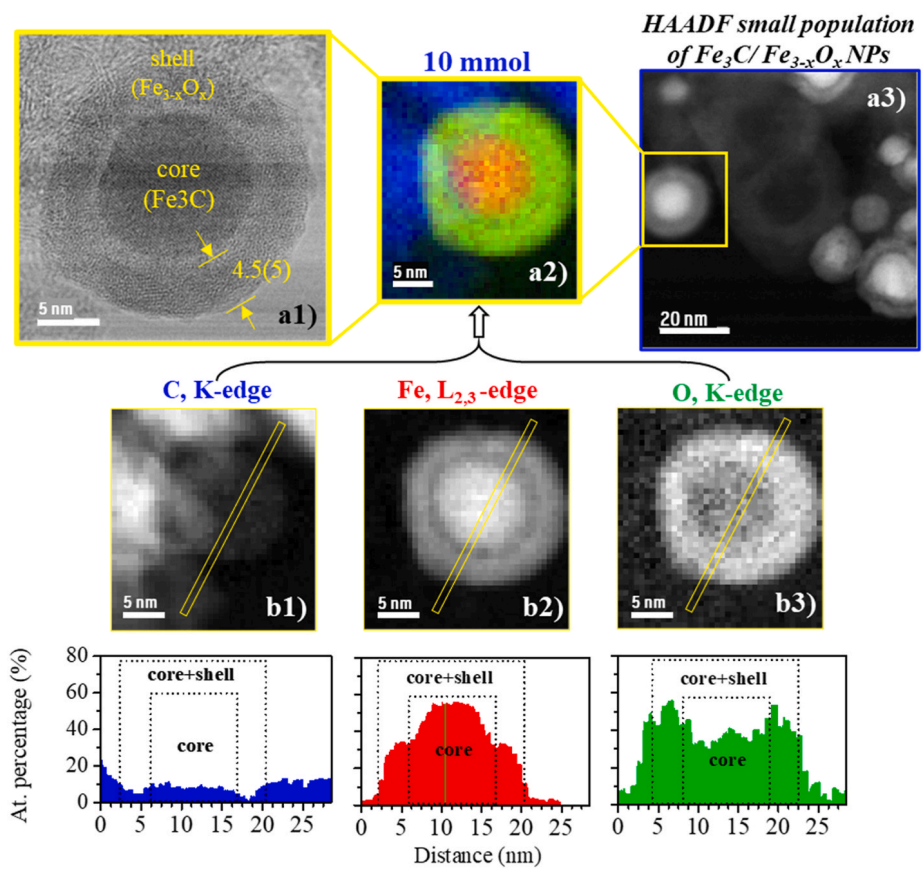

Fig. 5. (a1) ABF image of a small core/shell NP of the 10 mmol sample. (a2) Coloured image generated by the superposition of the elemental compositional maps obtained from the C $K$ (blue), Fe $L_{2,3}$ (red) and O $K$ (green) edges. (a3) HAADF image of some $Fe_3C/Fe_{3-x}O_x$ NPs. (b1-b3) Compositional maps (atomic percent) for the (a1) NP based on the analysis of the C and O $K$-edges, Fe $L_{2,3}$-edge. At the bottom of each histogram, the profile composition extracted from the regions marked with a yellow rectangle is shown. (For interpretation of the references to color in this figure legend, the reader is referred to the Web version of this article.)

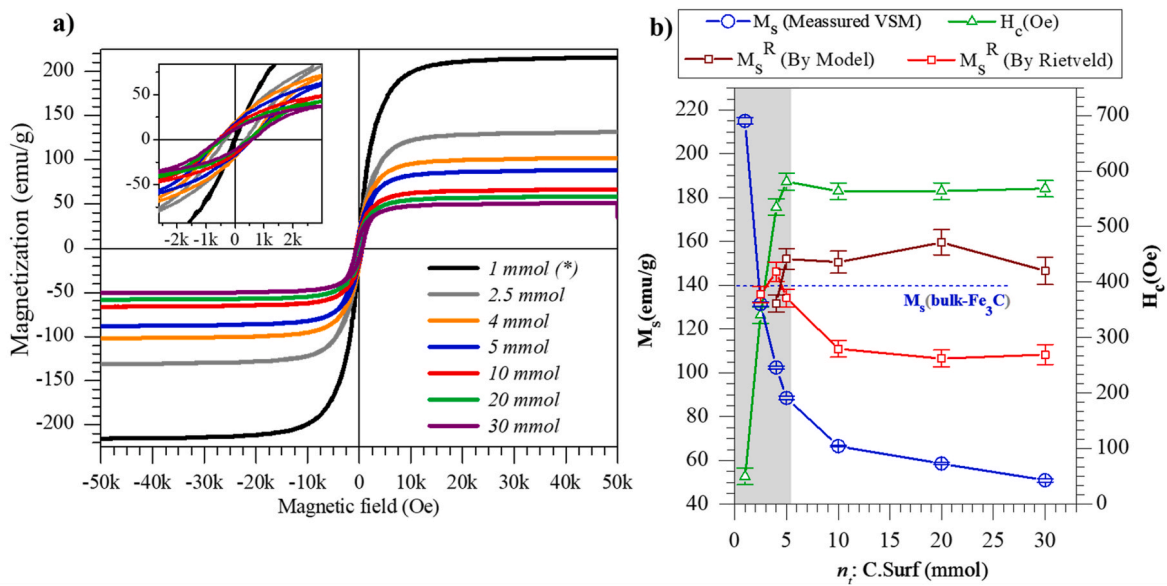

Fig. 6. (a) Magnetic hysteresis loops taken at 300 K for the samples, with 1 mmol (*) sample reprinted from Ref. [57]. (b) Saturation magnetization (blue), and coercive field (green) dependencies as a function of each surfactant concentration $n_t$. $M_s^R$ values with the weight percentages calculated by Rietveld (red) and carbon matrix predicted by the proposed model (brown). (For interpretation of the references to color in this figure legend, the reader is referred to the Web version of this article.)

Table 3
Summary of the magnetic parameters of the hysteresis loops for samples with different surfactant concentration ($n_t$, for each surfactant added), and an additional 1 mmol (*) sample.

| Sample | $M_s \pm 1$ (emu/g) | From VSM $M_s^R$ (emu/g) | From Model $M_s^R$ (emu/g) | $H_C \pm 15$ (Oe) |
|---|---|---|---|---|
| 1 mmol (*) | 215 | – | – | 39 |
| 2.5 mmol | 131 | 147 ± 6 | – | 340 |
| 4 mmol | 102 | 148 ± 5 | 132 ± 4 | 535 |
| 5 mmol | 86 | 133 ± 4 | 152 ± 5 | 580 |
| 10 mmol | 67 | 110 ± 3 | 150 ± 5 | 560 |
| 20 mmol | 59 | 106 ± 4 | 160 ± 6 | 565 |
| 30 mmol | 51 | 113 ± 5 | 146 ± 6 | 570 |

magnetic nature of α-Fe. On the other hand, with decreasing $n_t$, the $M_s$ values first show a linear increase from 30 to 10 mmol, correlated with the lessen C matrix. However, for $n_t < 10$ mmol, $M_s$ shows a sharp increase reaching values of 86(1), 102(1) and 131(1) emu/g for 5, 4 and 2.5 mmol respectively, until reaching 215 emu/g for 1 mmol (pure α-Fe). Therefore, considering both the $H_C$ and $M_s$ trends with $n_t$, it is concluded that there is a critical transition region at $n_t = 2.5$–4 mmol, where the carbon content is insufficient to reduce $Fe_2O_3$ to Fe and carburize it. It should be noted that the 2.5 and 4 mmol samples display a high degree of compositional heterogeneity (Fig. 3-b1), while the 5 mmol sample exhibits multiple PSDs (Fig. 4-a1). In summary, the structural and magnetic results indicate that there is a threshold carbon content in the xerogel that must be exceeded to fully carburize the iron.

As a further step, we have developed a model to elucidate how the reduction in the amount of surfactant introduced during the synthesis leads to this transition region. For this purpose, an approximative count of the mass loss of the xerogel during the heat treatment is proposed, which predicts the amount of residual carbon that would form the carbon matrix at the end of the heat treatment (see Supporting Information for more details). Taking into account the necessary considerations for the loss of $H_2O$, $NO_3^-$, and surfactants, combined with the reduction of Fe-oxides by the consumption of C via its oxidation ($C_{(s)} + 2O_{(\cdot)} \rightarrow CO_2$ (g)), the model predicts a threshold value, $(n_t)_{min}$, for the complete carburization of Fe-oxides into Fe3C of nt 3.1(6) mmol. The threshold is consistent with that shown in Fig. 3-b1), where the amount of carbon is insufficient to carburize Fe completely in 2.5 and 4 mmol samples. For the 2.5 mmol sample, the carbon content is practically zero and it defines the region where the magnetic properties begin to destabilize (Fig. 6-b). Nevertheless, it is noteworthy that the model provides a reliable indication of the limiting $n_t$ values, although it does not take into account experimental facts such as: i) the presence of residual $Fe_{3-x}O_x$ oxides after densification, and ii) the effects of the $N_2$ atmosphere, which may remove oxygen from the system and mediate by strongly reducing species formation of $C_yN_x$, which could decrease the $(n_t)_{min}$. Therefore, a perfect agreement of the $(n_t)_{min}$ calculation with the experimental values is not to be expected. However, the validity of the model was also be confirmed by reproducing the amount of carbon matrix remaining after the densification process (see Supporting *Information*). This quantity was used to obtain the $M_s^R$ values (Fig. 6-b). Such $M_s^R$ values (brown data) improve the infra-estimation of those obtained by Rietveld refinement (red data), consistent with the presence of possible amorphous carbon in the matrix not detected by XRD. Therefore, the suggested model has proved to be useful for predicting a preliminary limiting $n_t$ value directly related with the magnetic optimization of the synthesized samples and could be extrapolated to similar synthesis methods.

In conclusion, due to the homogeneity of the samples shown in terms of morphology (Fig. 4), phase composition (Fig. 3) and magnetic properties (Fig. 6), the samples selected for viability are those comprised between 5-30 mmol.

To put our results into context, a comprehensive review of the state of art on synthesis methods of Fe-carbides, and specifically $Fe_3C$ NPs is summarized in Table 4. Overall, $Fe_3C$ shows lower coercive fields ($100 < H_C < 300$ Oe) for larger NP sizes, regardless of the synthesis approach followed. In turn, the XRD-phase analysis of the examined compounds reveals that the existence of magnetically soft α-Fe impurities also contributes to the reduction of the $H_C$, in agreement with Fig. 6-b). Most sol-gel chemistry routes produced samples with NPs embedded in a carbon matrix and coated with a few layers of carbon. Studies such as [64] have evaluated the protective effect of carbon coatings on $Fe_3C$ in relation to oxidation stability. The authors report $M_s = 132$ emu/g for $Fe_3C$ samples and a lower $M_s = 95$ emu/g for α-Fe samples due to its high degree of oxidation. In addition, others have reported $M_s$ values above 100 emu/g along with $H_C \sim 500$-600 Oe, features typically attributed to Fe-carbide phases. In these works, the samples were not single-phase and $Fe_3O_4$ shells were identified because carbon passivated shells were not achieved [38]. Here we assess all these facts systematically in the synthesis methodology presented. Furthermore, among all the sol-gel synthetic routes reviewed, the samples presented in this work display one of the highest $M_s$ values reported for pure $Fe_3C$/C encapsulated core/shell NPs, along with a well-control of α-Fe impurities. Specifically, $M_s = 86$ emu/g for pure $Fe_3C$ at $n_t = 5$ mmol and 102 emu/g for $Fe_3C$ with an extremely small α-Fe impurity (2 wt%) at $n_t = 4$ mmol. However, to evaluate the cytotoxicity of the samples, we only restrict the assays to 5–30 mmol samples due to phase purity and PSD.

Table 4
Summary of the magnetic parameters and particle sizes of different nanoparticle systems of Fe-carbides obtained by different synthesis approaches.

| Article | $M_s \pm 1$ (emu/g) | $H_C \pm 5$ (Oe) | XRD-phases & Morphology | NP-size (nm) | C-matrix embedded | Synthesis route |
|---|---|---|---|---|---|---|
| [39] | 139 | 141 | $Fe_3C$ | 100–200 | no | colloidal |
| [41] | 48 | 75 | Fe-carbides | 100–200 | no | plasma discharge |
| [42] | 66 | 300 | $Fe_3C$ & α-Fe | 15–160 | no | nanocasting |
| [34] | 123 | – | $Fe_3C$/C core/shell | 25–35 | no | reducing flame spray pyrolysis |
| [20] | 88 | 175 | $Fe_3C$ | 20–35 | no | calcination |
| [64] | 132 | 560 | $Fe_3C$/C core-/shell & α-Fe | 20–30 | no | laser pyrolysis |
| [24] | 125 | – | $Fe_5C_2$/$Fe_3O_4$ core/shell | 10–15 | no | wet-chemistry-route |
| [38] | 101 \| 95 \| 88 | 545 \| 635 \| 655 | $Fe_3C$ \| $Fe_5C_2$ \| $Fe_2C$/$Fe_3O_4$ core/shell | 10-20/2-4 | no | colloidal |
| [40] | – | – | $Fe_5C_2$ or $Fe_2C$ or $Fe_7C_3$ | 15–20 | no | liquid-phase route |
| [45] | 138 | 205 | $Fe_3C$ & α-Fe | 100–200 | yes | sol-gel pyrolysis |
| [43] | 40 | 210 | $Fe_3C$ & α-Fe (~15 %) | 20–100 | yes | glucose-pyrolysis |
| [44] | 104 | 470 | $Fe_3C$ | 10–20 | yes | sol-gel pyrolysis |
| [10] | 63 | 220 | $Fe_3C$/C core/shell | 25–35 | yes | sol-gel pyrolysis |
| [11] | 24 | 550 | $Fe_3C$/C core/shell, & α-Fe | 20–50 | yes | metal-organic chemical vapor deposition |
| [13] | 51 | 440 | $Fe_3C$/C core/shell | 15–25 | yes | sol-gel pyrolysis |
| This work | 102–86 | 535–580 | $Fe_3C$/C core/shell | 15–45 | yes | sol-gel pyrolysis |

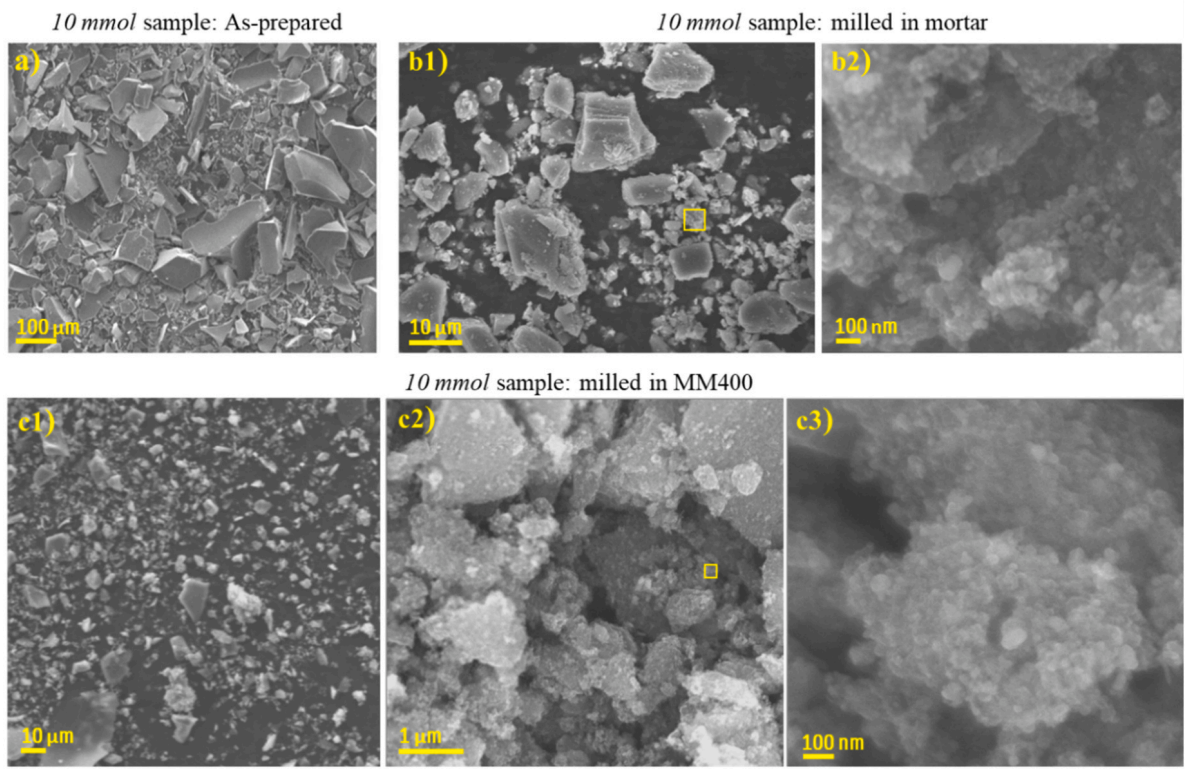

**Fig. 7.** SEM images for 10 mmol sample: (a) as-prepared powders and (b1) milled in agate mortar, (b2) zoom image from the squared region in (b1) and (c1-3) milled in MM400.

### 3.4. Cell viability studies

The 10 mmol sample represent an optimal balance between morphological (Fig. 4-a2) and compositional homogeneity (Fig. 3), while also exhibiting a high $M_s$ value. Consequently, it was chosen for cell viability studies. Prior to this, further particle size reduction and homogenization was carried out through a 40 min milling in a mixer mill MM400 (details can be found in the Supporting Information) and the optimized sample was labeled 10 mmol *MM400*. Fig. 7 shows the SEM images of as-prepared and agate mortar-milled 10 mmol sample for comparison. A large decrease in the powder particle size can be seen from the as-prepared sample (Fig. 7-a) to the mortar-milled ones (Fig. 7-b1-b2). In turn, a similar result is obtained for MM400-milled sample (Fig. 7-c1) and c2)), but with more homogeneous size distribution. It is worth noting that despite the slightly large size of the obtained powder particle (≤10 μm), even after milling (Fig. 7-b3) and c3)), compared to the cells dimensions, many spherical-like structures of nanometer size (∼50 nm or less) are observed. These particles will interact directly with the cells and therefore their specific surface will be determinant for the viability studies.

The cytotoxic impact of nanoparticles was assessed in MCF-7 cells (breast cancer cell line) using the Alamar colorimetric cell viability assay. Following a 24-h treatment with nanoparticles at different surfactant proportions (5, 30, 10 mmol (as prepared and milled)) and extracellular concentrations ranging from 12.5 to 100 μg/mL, the viability of MCF-7 cells showed no significant decrease (see Fig. 8), except for the samples with 5 mmol surfactant. This finding suggests a positive correlation between elevated surfactant content employed in the synthesis with higher viability values. In the same line, the viability values of the sample with sizes and morphologies optimized by mechanical milling (10 mmol milled) are slightly lower compared to those of the 10 mmol sample in its as-prepared state. This could indicate that an optimization of particle size and morphology promotes a better interaction with the cells. The results are interesting because even using a higher concentration of nanostructures (100 μg/mL), the viability is larger than 85 %. Consequently, it leads to the inference that the synthesized nanoparticles represent advantageous materials suitable for a range of magnetic-based biomedical applications.

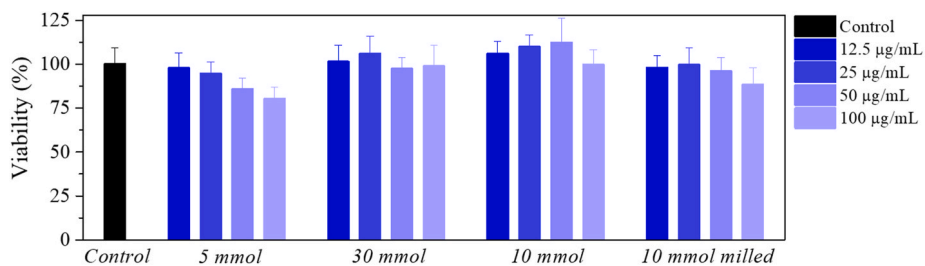

**Fig. 8.** Cell viability values of MCF-7 cells after incubating NPs from samples with different concentrations for each surfactant added (5, 30, 10 mmol, as prepared and milled) at 37 °C for 24 h at extracellular concentrations ranging from 0 to 100 μg/mL.

## 4. Conclusions

A novel sol-gel optimization for obtaining pure $Fe_3C/C$ encapsulated core/shell NPs supported by a carbon matrix is developed by reducing each surfactant concentration added ($n_t$). A combined analysis of the compositional crystallographic phases and the morphology of the particles leads to the existence of a limiting amount of surfactant, $(n_t)_{min}$ ~4 mmol. Below this value, the system exhibits α-Fe contamination, and the coalescence of NPs is promoted (even at $n_t \leq 5$ mmol) as the C-matrix is reduced. Consequently, a spreading of the mean particle size from 15-25 nm to 15–45 nm is observed (from *30* to 5 mmol samples). Although the presence of some $Fe_3C/Fe_{3-x}O_x$ core/shell NPs was systematically found, the low compositional percentage of Fe-oxide extracted from the Rietveld analysis compared to that of graphite indicates that the samples are mostly constituted of NPs with a $Fe_3C$ core and a well-defined multilayered graphene/graphite shell. The magnetic characterization showed that decreasing $n_t$ modulates the magnetic response of the samples, achieving $M_s$ around 86–102 emu/g for 5 and 4 mmol samples respectively. $H_C$ remains constant around 550 Oe above $(n_t)_{min}$ value in agreement with single phase $Fe_3C$ samples, whereas it falls below $(n_t)_{min}$ because of the soft magnetic behaviour of α-Fe. A model was proposed to evaluate the reduction of the surfactant and its corresponding carbon matrix content, predicting a $(n_t)_{min}$ in agreement with the experimentally observed. This model is crucial to optimize the magnetic response of the composites in a predictive way. A review of the state of the art has been carried out showing that the samples reported here have competitive magnetic properties combined with high purity compared to other synthetic approaches. Finally, biocompatibility cell assays on different samples showed a remarkably low cytotoxicity of the NPs on breast cancer cells (MCF-7 cell line), reflecting a potential applicability in biomedical fields, particularly those associated with magnetic-based theragnostic treatments

## CRediT authorship contribution statement

**A. Castellano-Soria:** Writing – review & editing, Writing – original draft, Validation, Methodology, Investigation, Formal analysis, Data curation, Conceptualization. **R. López-Méndez:** Writing – review & editing, Data curation. **A. Espinosa:** Writing – review & editing, Validation, Supervision, Investigation, Formal analysis. **C. Granados-Miralles:** Writing – review & editing, Resources, Investigation. **M. Varela:** Writing – review & editing, Visualization, Validation, Resources, Investigation, Formal analysis. **P. Marín:** Writing – review & editing, Validation, Supervision, Resources, Funding acquisition, Conceptualization. **E. Navarro:** Writing – review & editing, Supervision, Resources, Methodology, Investigation, Conceptualization. **J. López-Sánchez:** Writing – review & editing, Validation, Supervision, Resources, Methodology, Investigation, Funding acquisition, Conceptualization.

## Declaration of competing interest

The authors declare that they have no known competing financial interests or personal relationships that could have appeared to influence the work reported in this paper.

## Data availability

Data will be made available on request.

## Acknowledgments

The work has been supported by the Ministerio de Ciencia e Innovación (MCINN) through the projects PDC2022-133039-I00, PID2021-123112OB-C21-MICIIN, TED2021-129688B-C21, MAT2015-65445-C2-1-R, MAT2017-86450-C4-1-R, MAT2015-67557-C2-1-P, RTI2018-095856-B-C21, RTI2018-095303-A-C52, PID2020-114192RB-C41, PIE: 2021-60-E-030, PIE: 2010-6-OE-013, PID2021-122980OB-C51, FPI PRE2020-96246 grant (R.L.-M.), and Comunidad de Madrid by NANOMAGCOST-CM, S2013/MIT-2850 NANOFRONTMAG, (MAD2D-CM)-UCM3, S2018/NMT-4321 NANOMAGCOST, 2018-T1/IND-1005 (A.E.) and S2022/BMD-7434 ASAP-CM. C.G.-M. acknowledges financial support from grant RYC2021–031181-I funded by MCIN/AEI/10.13039/501100011033 and by the "European Union NextGenerationEU/PRTR". J. L.-S. acknowledges the financial support from Spanish Ministry of Economic Affairs and Digital Transformation through the project PID2021-122980OB-C5 and from grant RYC2022-035912-I funded by MCIU/AEI/10.13039/501100011033 and by the European Social Fund Plus (ESF+). Electron microscopy measurements were carried out at the Centro Nacional de Microscopía Electrónica at the Universidad Complutense de Madrid (ICTS ELECMI, UCM). The authors also acknowledge the BM25-SpLine Staff for their valuable help and the provision of the beamline under the projects A25-2-1053 and A25-2-1009.

## Appendix A. Supplementary data

Supplementary data to this article can be found online at https://doi.org/10.1016/j.mtchem.2024.102143.